\documentstyle[preprint,aps]{revtex}
\begin{document}
\draft
\title{\bf{\LARGE{Localized states due to the coupling of exciton with
the coupled lattice oscillators}}}
\author{Bikash Chandra Gupta}
\address{Department of Physics, Indian Institute of Technology-Madras,\\ 
Chennai 600036, India}
\maketitle
\begin{abstract}

Discrete nonlinear Schr\"oginger equation (DNLS) of the form, $i \frac{dC_n}  
{dt}$ = $C_{n+1}$ + $C_{n-1}$ - $ \chi_n [|C_{n+1}|^2 + |C_{n-1}|^2 - 
2 |C_n|^2] C_n$ is used to study the formation of stationary localized
states in one dimensional system due to a single as well as a dimeric
nonlinear impurity. The fully nonlinear chain is also
considered.  The stability of the states and its connection with the 
nonlinear strength is presented. Results are compared with those obtained 
from other DNLS. It is found that the DNLS used in this paper has more 
impact in the formation of stationary localized states.

\end{abstract}

\narrowtext

\section{Introduction}

It is well know that the transport properties of a system is directly related 
with the formation of localized states in the system. Localized states 
appear due to the presence of impurity or disorder (which breaks the 
translational symmetry) in the system \cite{econ}. 
There has been studies on the 
formation of localized states due to linear impurities in various systems
\cite{econ}. 
On the other hand, only a few look into the formation of localized states 
due to nonlinear impurities. The discrete nonlinear Schr\"odinger equation, 
used to study the formation of stationary localized (SL) states 
\cite{mol1,mol2,mol3,hui1,hui2,acev,wein,bik2,bik3,kund,bik1,ghos} is 
given by  
\begin{equation}
i\frac{dC_n}{dt}=\epsilon_n C_n + V(C_{n+1} + C_{n-1}) 
- \chi_n |C_n|^{\sigma} C_n,
\end{equation}
here $C_n$ is the probability amplitude of the particle (exciton) to be
at the site $n$, $\epsilon_n$ and $\chi_n$ are the static site energy 
and the nonlinear strength at site $n$ respectively and $V$ is the 
nearest neighbor hopping element. The nonlinear term $|C_n|^\sigma C_n$
arises due to the interaction of the exciton with the lattice 
vibration \cite{ken1,ken2}. The above eq. (1), has been used to study 
the formation of SL states in one dimensional chain as well as in Cayley 
tree with single and dimeric nonlinear impurities
\cite{mol1,mol2,mol3,hui1,hui2,bik2,bik3,kund}. For the case of a
perfectly nonlinear chain where $\chi_n=\chi$, it has been shown that 
SL states are possible even though the translational symmetry of
the system is preserved \cite{acev,wein,kund,bik1,ghos}. 
These results were remarkably different when
compared with that of the corresponding systems with linear impurities. 

The equation (1) has been derived with the assumption 
that the lattice oscillators in the system are local and oscillate
independently. A natural question to ask is, what happens to the formation of
SL states when the oscillators in the lattice are occupied with their nearest
neighbors. In this case the discrete nonlinear Schr\"odinger 
equation takes the form 
\begin{equation}
i\frac{dC_n}{dt}=\epsilon_n C_n + V(C_{n+1} + C_{n-1}) - \chi_n (|C_{n+1}|^2
+|C_{n-1}|^2 - 2|C_n|^2)C_n
\end{equation}
where $C_n$, $\epsilon_n$, $V$ and $\chi_n$ carries the same meaning as 
in eq. (1). We notice that the eq. (2) has more nonlinear terms compared 
to the eq. (1). To the best of our knowledge, this equation has not been
used to study the formation of SL states even though eq. (2) is
more important in condensed matter physics. Our intention  is to look
for the formation of SL states in one dimensional system due to the
presence of nonlinear impurities (as described by eq. (2)) and further 
to compare the results  with those obtained from eq. (1) and to see which 
one has more impact in the formation of SL states.    

The organization of the paper is as follows. In sec. II we discuss the
effect on the formation of SL states due to a single impurity ({i.e.}
$\chi_n=\chi(\delta_{n,0})$). In sec. III we consider the case of 
dimeric impurity ({\em i.e.} $\chi_n = \chi (\delta_{n,0} + \delta_{n,1})$) 
and in sec IV we consider the perfectly nonlinear chain. In 
sec V we discuss about the stability of the SL states. Finally in sec. VI
we summarize our findings.

\section{Single Nonlinear Impurity}

Consider the system of a one dimensional chain with a nonlinear impurity 
at the central site. The time evolution of an exciton in the system is 
governed by eq.(2) with $\chi_n = \chi \delta_{n,0}$. The Hamiltonian
which can produce the equation of motion for the exciton in the system 
is given by
\begin{equation}
H=\sum_n (C_n^\star C_{n+1} + C_n C_{n+1}^\star) - \frac{\chi}{2} [|C_1|^2
+ |C_{-1}|^2 - 2 |C_0|^2] |C_0|^2.
\end{equation} 
As $\sum_n |C_n|^2$ is a constant of motion, we suitably renormalized so that
$\sum_n |C_n|^2$=1. We call it normalization constant. Therefore, $|C_n|^2$ 
can be treated as the probability for the exciton to be at site $n$. Since 
we are interested in finding the stationary localized states, we consider
the ansatz
\begin{equation}
C_n=\phi_n exp(-iEt);~~~~~\phi_n=\phi_0 \eta^{|n|}
\end{equation}
where $0 < \eta <1$ and $\eta$ can be asymptotically defined as $\eta =
\frac{|E|-\sqrt{E^2-4}}{2}$. $E$ is the energy of the localized state which
appears outside the host band. Since in a one dimensional system, states
appearing out side the host band are exponentially localized, the ansatz 
(in eq. (4)) is justified as can also be readily derived from the Greens 
function analysis \cite{bik2,bik3}. Substituting the ansatz in the 
normalization condition we get
\begin{equation}
|\phi_0|^2 = \frac{1-\eta^2}{1+\eta^2}.
\end{equation}
Direct substitution for $\phi_0$, $\phi_n$ and hence $C_n$ in terms of
$\eta$ in eq. (3), yields an effective Hamiltonian,
\begin{equation}
H_{eff} = \frac{4\eta}{1+\eta^2} + \frac{\chi (1-\eta^2)^3}{(1+\eta^2)^2}.
\end{equation}
The fixed point solutions of the reduced dynamical system described by 
$H_{eff}$ will give the values of $\eta$ (which correspond to 
the localized state solutions) \cite{wein}. Note that 
the effective Hamiltonian is a function of 
only one dynamical variable, namely, $\eta$ as $\chi$ is constant. 
Thus fixed point solutions are readily obtained from the condition
$\partial{H_{eff}} /\partial{\eta}$ = 0, {\em i.e.},
\begin{equation}
\frac{4}{\chi}=\frac{\eta (1-\eta^2) (10+2\eta^2)}{(1+\eta^2)} =f(\eta).
\end{equation}
Thus the different values of $\eta \in$ [0,1] satisfying the eq. (7) will 
give the possible SL states for a given value of $\chi$. It is clear from 
the expression for $f(\eta)$ that $f(\eta)\rightarrow 0$ as $\eta\rightarrow 
0$ and $\eta \rightarrow 1$.
Therefore it is expected that $f(\eta)$ will have at least one maximum, 
which is indeed the case as can be seen from Fig. (1) where $f(\eta)$ 
is plotted as a function of $\eta$.
Notice that there will be no graphical solution if $\frac{4}{\chi} >
f(\eta_{max})$, one solution if $\frac{4}{\chi}=f(\eta_{max})$ and two
solutions if $\frac{4}{\chi}<f(\eta_{max})$. Thus there is a critical
value of $\chi$, say, $\chi_{cr}$ below which no localized states are 
possible and  is given by
\begin{equation}
\chi_{cr}=\frac{\eta_{max} (1-\eta_{max}^2) (10 + 2\eta_{max}^2)}{4
(1+\eta_{max}^2) } = 1.2696.
\end{equation}     
Thus for $\chi$=1.2696, we get one SL state and two for $\chi >$ 1.2696. 
For a system described by eq. (1) (with $\sigma$=2), it has been shown in ref.\cite{mol2,bik2}
that the corresponding critical value for $\chi$ is 2. Also the maximum 
number of SL states possible was 1. Thus we see that the nonlinearity 
arising in eq. (2) reduces the critical strength and produces more number 
of SL states. Hence, the eq. (2) is indeed more effective in the formation 
of SL states.

\section{Dimeric Nonlinear Impurity}

We consider the case where the one dimensional lattice has two nonlinear
impurities at site 0 and 1 respectively, {\em i.e.}, $\chi (\delta_{n,0} 
+ \delta_{n,1})$. As in the case of single impurity, it is easily verified 
that the Hamiltonian for the system is given by eq. (2) with $\chi$ as 
defined above. For stationarity condition we assume that
$C_n = \phi_x exp(-iEt)$. Furthermore, for localized states we assume
the following form for $\phi_n$.
\begin{eqnarray}
\phi_n = [sgn(E) \eta]^{n-1} \phi_1 ; ~~~~~~~~~~~~~~n \ge 1 \nonumber \\
 {\rm and} \nonumber \\
\phi_{-|n|} = [sgn(E) \eta]^{|n|} \phi_0 ; ~~~~~~~~~~~~n \le 0
\end{eqnarray}
with $\eta$ as defined earlier. The ansatz is justified as those
states which appear outside the host band are exponentially localized 
(which can be derived exactly from the Greens function analysis 
\cite{bik2}). Three different possibilities arise.
(i) $\phi_1 = \phi_0$ (symmetric case), (ii) $\phi_1 = -\phi_0$ 
(antisymmetric case) and (iii) $\phi_1 \ne \phi_0$ (asymmetric case). 
It is possible to encompass all the different cases by introducing a 
variable $\beta = \frac{\phi_0}{\phi_1}$. The value of $\beta$ is 
confined between 1 and -1 if $|\phi_0| \le \phi_1$. Else we inverse 
the definition of $\beta$. $\beta = 1, -1, {\rm and} \ne 1$ correspond 
to the symmetric, antisymmetric and the asymmetric state respectively. 
Substituting the ansatz as well as the definition of $\beta$ in the 
normalization condition, $\sum_{-\infty}^{\infty}|C_n|^2 = 1$ we get
\begin{equation}
|\phi_0|^2 = \frac{1-\eta^2}{1+\beta^2}
\end{equation}  
and the reduced Hamiltonian 
\begin{equation}
H_{eff} = 2 \beta \frac{1-\eta^2}{1+\beta^2} + 2 sgn(E) \eta -
\frac{\chi (1-\eta^2)^3}{2 (1+\beta^2)} .
\end{equation}
If $\beta = \pm 1$ we get, 
\begin{equation}
H_{eff}^{\pm} = \mp 2\eta + 2 sgn(E) + 3 \chi_{\pm} 
\eta \frac{(1-\eta^2)^2}{2} 
\end{equation}
Here '+' sign corresponds to the symmetric case and '-' sign corresponds to
the antisymmetric case. The number of  fixed point solutions of the reduced
dynamical system described by $H_{eff}$ gives the the possible number of SL 
states. The fixed point solutions satisfy the equation,   
\begin{equation}
\frac{1}{\chi_{\pm}} = 3 \eta (1 \mp \eta) (1 \pm \eta^2). 
\end{equation}
From eq.(13) it is clear that there exists two critical values of
$\chi$ namely, 0.7149 and 1.6525. There is no SL state for $\chi < 0.7149$,
one symmetric SL state at $\chi$ = 0.7149, two symmetric SL states for $0.7149
< \chi <1.6525$, two symmetric and one antisymmetric SL state at $\chi$ =
1.6525 and two symmetric and two antisymmetric SL states for $\chi > 1.6525$. 

Now let us consider asymmetric case where $\beta \ne 1$. The
effective Hamiltonian is function of two dynamical variables namely,
$\beta$ and $\eta$. Therefore the fixed point solutions will obey the
equations given by 
\begin{equation}
\frac{\partial{H_eff}}{\partial\eta} = 0 ~~{\rm and} ~~\frac{\partial
H_{eff}}{\partial{\beta}} = 0.
\end{equation} 
After a little algebra we obtain the desired equation,
\begin{equation}
\frac{1}{\chi} = \frac{\beta (9 - 7 \beta^2 - \beta^4 - \beta^6)^2} {2
(1-\beta^2) (3-\beta^2)^4} = f(\beta).
\end{equation}
The function $f(\beta)$ monotonically increases with $\beta$ and it goes 
to infinity as $\beta$ goes to 1. From this we can immediately see that 
there always exists one SL state no matter how small $\chi$ may be. 

Combining all the possible states we find that there is one SL state for
$\chi < 0.7149$, two  at $\chi = 0.7149$, three  for
$0.7149 < \chi < 1.652$, four at $\chi = 1.6525$ and five 
for $\chi > 1.6525$. Hence the maximum number of SL states is five.
We further note that the critical values for nonlinear strength is lower 
and the number of SL states are more compared to the results
obtained by eq. (1) (with $\sigma$=2) \cite{bik3}. Thus it is again 
confirmed that eq. (2) 
is more effective in the formation of SL states compared to eq. (1).

\section{Fully Nonlinear Chain}

We now consider perfectly nonlinear chain {\em i.e.}, $\chi_n=\chi$. The
Hamiltonian for this system is given by eq.(2) with $\chi_n = \chi$. 
Using the stationarity condition, we can
obtain the Hamiltonian in terms of $\phi_n$. In this case it is not possible 
to find the exact
ansatz for the localized states, but there are a few rational choices.
For example, a single site peaked as well as inter-site peaked and
dipped solutions are possible. We will consider these cases
subsequently. Let us first consider the on-site peaked solution. Without
any loss of generality we can assume that the exciton profile is peaked at
the central site. Therefore, using the ansatz $\phi_n = \phi_0 \eta^{|n|}$ 
and the normalization condition we get the effective Hamiltonian, 
\begin{equation}
H_{eff}=\frac{4\eta}{1+\eta^2} + \chi \frac{(1-\eta^2)^3}{(1+\eta^2)^3}.
\end{equation}

From the fixed points equation, $\partial{H_{eff}}/\partial{\eta}$ = 0, 
we obtain
\begin{equation}
\frac{1}{\chi} = \frac{3\eta(1-\eta^2)}{(1+\eta^2)^2}.
\end{equation}

After analyzing this equation we find that there is a critical value of $\chi$ 
= 1.333 below which there is no SL state and
above it there are two states. At the critical value of $\chi$ there is
one state.

For the inter-site peaked and dipped solutions we use the ansatz
of the dimeric impurity nonlinear impurity. carrying out the calculation
involved we obtain the effective Hamiltonian of the reduced dynamical
system to be
\begin{equation}
H_{eff} = 2\beta \frac{1-\eta^2}{1+\beta^2} + 2 sgn(E) \eta -
\chi \frac{(1-\eta^2)^2}{(1+\beta^2)^2} [\beta^2 + \frac{\eta^2 + \beta^2
\eta^2 - 1-\beta^4}{1-\eta^4}]
\end{equation}
where $\beta$ is defined earlier.
We first consider the case
$\beta = \pm 1$. Substituting $\beta = \pm 1$ into the Hamiltonian
and from the fixed points equations we obtain 
\begin{equation}
\frac{1}{\chi_{\pm}}= \frac{\eta (1-\eta^2)^2 (2+\eta^2)}{2 (sgn(E) \mp \eta) 
(1+\eta^2)^2}.
\end{equation}
Here '+' sign corresponds to the symmetric case and the '-' sing 
to the antisymmetric case.
From eq. (19) it is clear that there will be two critical values of $\chi$
namely, $\chi_{cr}^+ = 2.4653$ and $\chi_{cr}^- = 5.9178$. There is no SL state
for $\chi < \chi_{cr}^+$, one SL state for $\chi = \chi_{cr}^+$, two SL
states for $\chi_{cr}^+ < \chi < \chi_{cr}^-$, three SL states at $\chi =
\chi_{cr}^-$ and four SL states for $\chi > \chi_{cr}^-$.

On the other hand for $\beta \ne 1$ we find that
$\beta \in [0,1]$ and $\eta \in [0,1]$ satisfy the 
following equations.
\begin{eqnarray}
-4\beta\eta (1+\beta^2) (1+\eta^2)^2 + 2 sgn(E) (1+\beta^2)^2 (1+\eta^2)^2
+ 2 \chi \eta [-3 + \beta^2 -\beta^2 \eta^4] \nonumber \\ 
- \chi [2 \eta^2 - 4 \beta^2 \eta^2 
-2 \beta^4 + \eta^4 - 2 \beta^2 \eta^6] = 0 \nonumber \\
2 (1+\eta^2) (1+\beta^2)^2 - 4 \beta^2 (1+\eta^2) (1+\beta^2) -
\chi [(1+\beta^2) (2 \beta - 2 \beta \eta^4 _ 2 \beta \eta^2 - 4 \beta^3] 
\nonumber \\
+4 \chi \beta [\beta^2 - \beta^2 \eta^4 + \eta^2 + \beta^2 \eta^2 -1 -\beta^4]
=0
\end{eqnarray}
As it is not possible to decouple the equations,  we have obtained
numerically the possible values of $\beta$ and $\eta$ for various values of 
$\chi$. It is found that there is always exists one SL state for any nonzero 
value of $\chi$.
  
Now combining all the possibilities, we obtain the following result for
the fully nonlinear chain. There will be only one SL states for $\chi <
2.4653$, two  for $\chi = 2.4653$, three  for $2.4653 < \chi < 5.9178$, 
four  for $\chi = 5.9178$ and five for 
$\chi > 5.9178$. Hence the maximum number of SL states is five. 
We further note that SL state appears even if the system is perfect 
(the translational symmetry is preserved). Therefore, we may call these 
states to be {\it self-localized} states.

\section{Stability}

The stability of the SL states can be understood from a  simple graphical 
analysis. For this purpose, consider the case of single impurity with
$\chi = 1.3$ (for which two SL states appear). The fixed point equation 
for the single impurity case with $\chi = 1.3$ is given by
\begin{equation}
G(\eta)= 1 - \frac{\eta (1-\eta^2) (10 + 2 \eta^2)}{(1+\eta^2)} = 0.
\end{equation}
The flow diagram of the dynamical system described by the $H_{eff}$ 
given in eq. (6) is constructed in the following manner. We treat $G(\eta)$
as the velocity and $\eta$ as the coordinate of the dynamical system.
$G(\eta)$ is plotted as a function of $\eta$ in fig. (2).  'A' and 'B' are 
the fixed points corresponding to the SL states. If $G(\eta) > 0$, the 
flow of the dynamical variable is in right direction else it is in 
left. The direction of flow is shown by arrows in 
different regions. It is clear from the flow diagram 
'A' is a stable fixed point whereas 'B' is unstable. 
Therefore, for the case of single impurity, one state is stable and the 
other one is unstable. 

The energy of the SL states as a function of $\chi$ is plotted in fig. (3).
Once again we confine to the single impurity case. It is clear that energy of
one state increases and that of the other decreases. (Note that the points 
'A' and 'B' of fig.(3) gets mapped to points '$A^\prime$' and '$B^\prime$'
respectively.) Thus we conclude that the states in the upper branch of 
the energy diagram are stable SL states and those of the lower branch are 
unstable SL states. In other words, if the energy of SL state increases 
with the increase of nonlinear strength, the state is stable otherwise,
unstable.

\section{conclusion}

DNLS given by eq. (2)
is used to study the formation of stationary localized states 
in a one dimensional system with single and a dimeric
nonlinear impurity. It is found
that the number of SL states are more than the number of impurities in the
system. Maximum number of SL states due to single nonlinear impurity is
two and that due to dimeric nonlinear impurity is five. It is further 
found that SL states may appear even in a perfectly nonlinear system. 
Thus one may call these SL states as {\it self-localized}
states. It is also interesting to note that eq. (2) is more effective in
the formation of SL states compared to eq. (1). The stability of the SL 
states is discussed
and the connection of the stability of a state with its energy variation
as a function of nonlinear strength is presented. For a clearer
understanding on the effect of nonlinear impurities on the formation of SL
states, one needs to consider the presence of a finite nonlinear clusters
in a linear host lattice. Investigation in this aspect is in progress and
will be reported elsewhere.

\section{Acknowledgement}
The author acknowledges the help from S. Seshadri during the preparation 
of the manuscript and the financial support from the Department of Science 
and Technology, India.

\begin{figure}
\caption{ $f(\eta)$ is plotted as a function of $\eta$. It clearly shows 
a maximum.}
\caption{G($\eta$) is plotted as a function of $\eta$. $A$ and $B$ are 
fixed points.}
\caption{Energies of the SL states are plotted as a function of $\chi$. 
There is no state below $\chi_{cr}$ and two states above it. Energy of 
one state increases and that of the other decreases. Points $A^\prime$ 
and $B^\prime$ corresponds to points $A$ and $B$ respectively of Fig. 2.}
\end{figure}


\begin{thebibliography}{99}
\bibitem{econ} E. N. Economou, {\it Green's Function in Quantum Physics}
               (Springer-Verlag, Berlin. 1979).
\bibitem{mol1} M. I. Molina and G. P. Tsironis, Phys. Rev. B {\bf 47} 15330
               (1993).
\bibitem{mol2} M. I. Molina, G. P. Tsironis and D. Hennig, Phys. Rev. E
               {\bf 50} 2365 (1994).
\bibitem{mol3} M. I. Molina and G. P. Tsironis, Int. Jour. Mod. Phys. B
               {\bf 9} 1899 (1995).
\bibitem{hui1} Y. Y. Yiu, K. M. Ng and P. M. Hui, Phys. Lett. A {\bf 200}
               325 (1995); Solid State Commun. {\bf 95} 801 (1995).
\bibitem{hui2} P. M. Hui, Y. F. Woo and W. Deng, J. Phys. Condens. Matt.
               {\bf 8} 2011 (1996).
\bibitem{acev} A. B. Aceves et al., Phys. Rev. E {\bf 53} 1172 (1996).
\bibitem{wein} B. Melomed and M. I. Weinstein, Phys. Lett. A {\bf 220}
               91 (1996).
\bibitem{bik2} B. C. Gupta and K. Kundu, Phys. Rev B {\bf 55} 894 (1997).
\bibitem{bik3} B. C. Gupta and K. Kundu, Phys. Rev B {\bf 55} 11033 (1997).
\bibitem{kund} K. Kundu and B. C. Gupta, Euro. Phys. Jour. B {\bf 3} (1998).
\bibitem{bik1} B. C. Gupta and K. Kundu, Phys. Lett. A {\bf 235} 176 (1997).
\bibitem{ghos} A. Ghosh, B. C. Gupta and K. Kundu, J. Phys. Conden. Matt.
               {\bf 10} 2701 (1998).
\bibitem{ken1} V. M. Kenkre, G. P. Tsironis and D. K. Campbell, {\it
               Nonlinearity in Condensed Matter}, eds. A. R. Bishop et al.
               (Springer-Verlag, 1987).
\bibitem{ken2} V. M. Kenkre and G. P. Tsironis, Phys. Rev. B {\bf 35}
               1473 (1987).
\end{thebibliography}
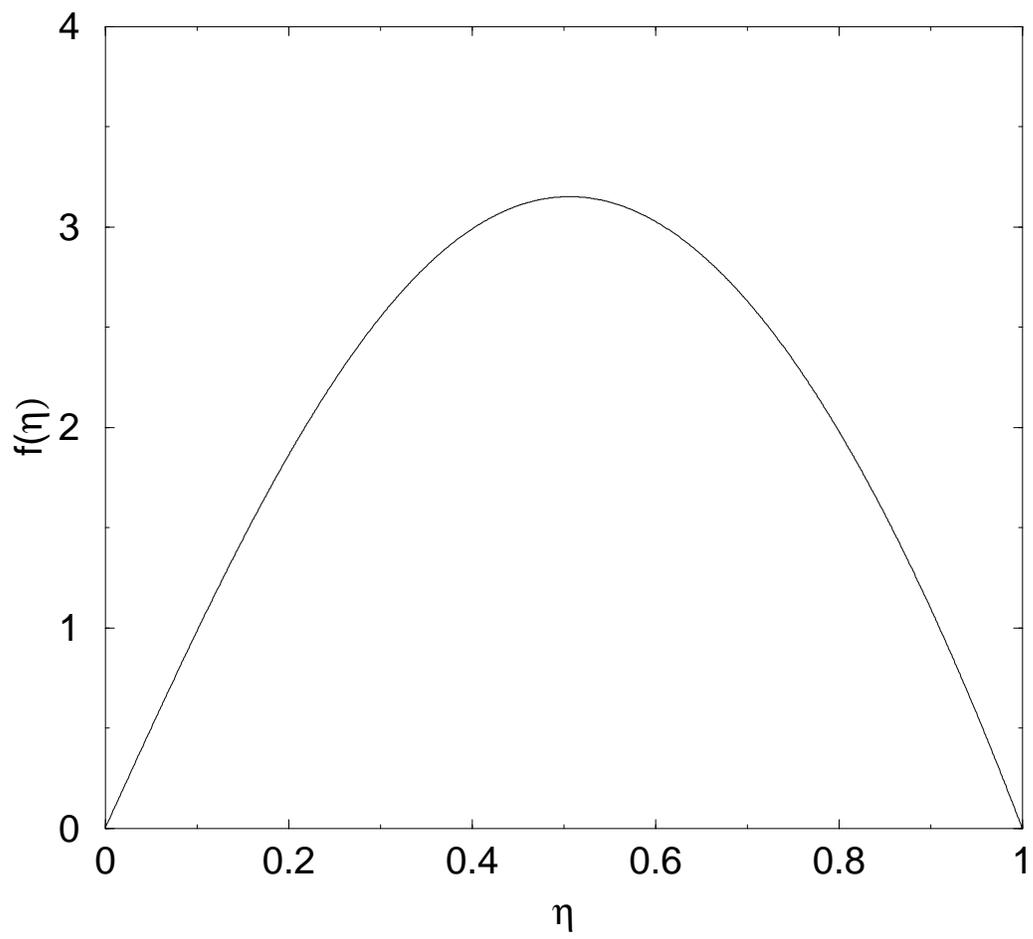
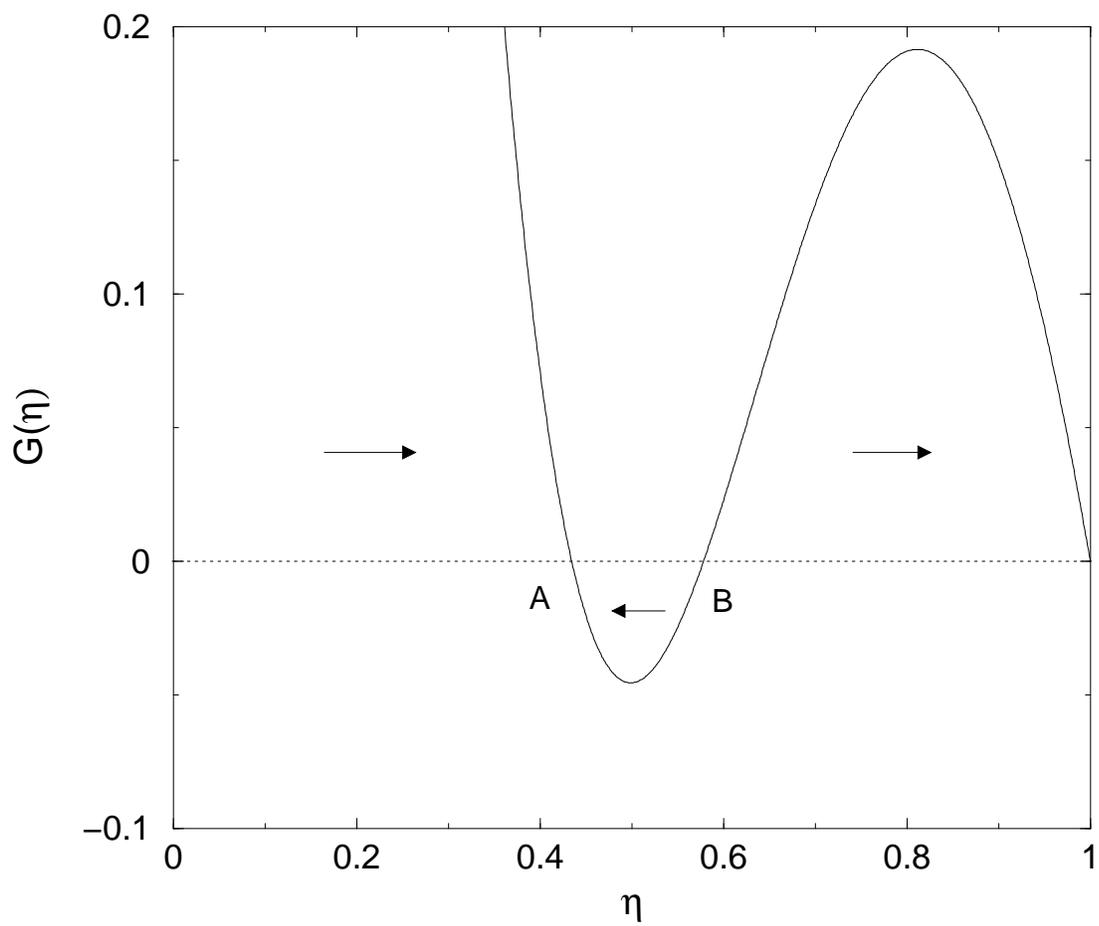
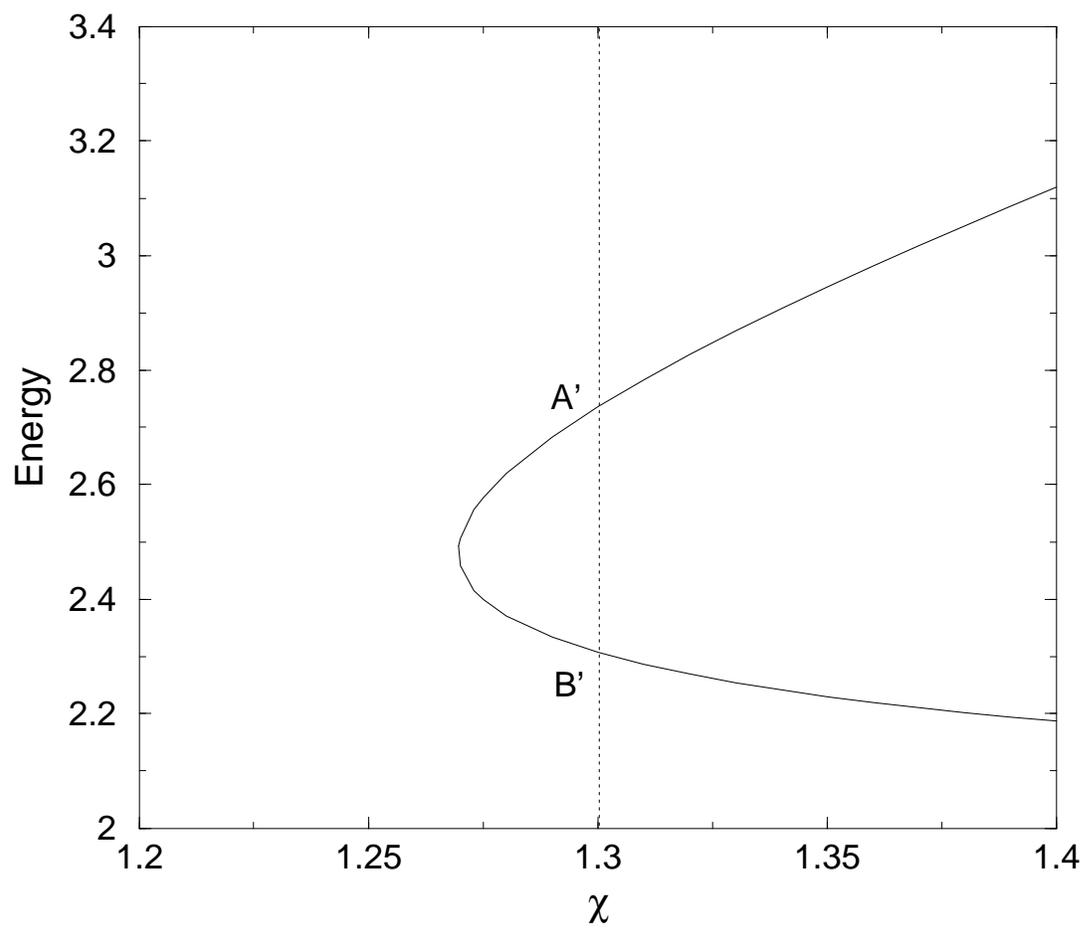
\end{document}